\documentclass[11pt,a4paper,onecolumn]{article}

\usepackage[utf8]{inputenc}
\usepackage[T1]{fontenc}
\usepackage[margin=1in]{geometry}
\usepackage{authblk}
\usepackage{cite}
\usepackage{amsmath,amssymb,amsfonts}
\usepackage{algorithmic}
\usepackage{graphicx}
\usepackage{textcomp}
\usepackage{comment}
\usepackage{xcolor}
\usepackage[normalem]{ulem}
\usepackage{hyperref}
\hypersetup{
    colorlinks=true, % убирает некрасивые рамки вокруг ссылок
    linkcolor=blue,  % цвет ссылок на рисунки, таблицы и разделы (\ref)
    citecolor=blue,  % цвет ссылок на список литературы (\cite)
    urlcolor=blue    % цвет ссылок на веб-сайты
}

\graphicspath{{./Bilder/}}

\title{Light-induced local control of magnetic anisotropy and domain pattern in Co/Pt multilayers}

\author[1,*]{Ivan Soldatov}
\author[1]{Aman Singh}
\author[1]{Volker Neu}
\author[1,2]{Rudolf Sch\"{a}fer}
\author[1]{Andrei Sherstobitov}
\author[1,2,3]{Yana Vaynzof}
\author[1,*]{Stanislav Bodnar}

\affil[1]{Leibniz Institute for Solid State and Materials Research (IFW) Dresden, Helmholtzstrasse 20, D-01069 Dresden, Germany}
\affil[2]{Institute for Materials Science, TU Dresden, 01062 Dresden, Germany}
\affil[3]{Chair for Emerging Electronic Technologies, Dresden University of Technology, Nöthnitzer Str. 61, 01187 Dresden, Germany}
\affil[*]{Corresponding authors: i.soldatov@ifw-dresden.de, s.bodnar@ifw-dresden.de}

\date{}

\begin{document}

\maketitle

\begin{abstract}
Control of magnetic anisotropy is essential for the fabrication of next-generation devices in spintronics and magnetic memory.
While conventional approaches such as material composition, thermal annealing, and ion bombardment are widely used, achieving spatially resolved, continuous control remains a challenge.
Herein, we demonstrate a dual-regime, localized control over both transient magnetic states and permanent intrinsic properties in perpendicularly magnetized Co/Pt multilayers using ultra-short laser pulses.
By employing a multi-scale approach combining Kerr microscopy, magnetic force microscopy, and vibrating sample magnetometry, we reveal that low-fluence irradiation enables a reversible, helicity-dependent accumulation of magnetic phases without altering the intrinsic quasistatic properties.
Conversely, high-fluence exposure acts as localized thermal annealing that permanently hardens the magnetic system.
This structural relaxation leads to a substantial enhancement of the effective perpendicular magnetic anisotropy, driving a twofold increase in the local coercive field and expanding the equilibrium stripe domain width.
By offering both reversible spin manipulation and deterministic engineering of 2D magnetic landscapes, this work paves the way towards flexible, lithography-free anisotropy structuring for advanced spintronic applications.
\end{abstract}

%\keywords{Perpendicular magnetic anisotropy, Co/Pt multilayers, Laser-induced modification, Magnetic domains, Magneto-optical Kerr microscopy, Magnetic force microscopy.}

\section{Introduction}
\label{sec:introduction}

One of the aims of spintronics is the control and readout of spin states in various magnetic systems such as ferromagnets, antiferromagnets and altermagnets \cite{bader2010spintronics, jungwirth2016antiferromagnetic, PhysRevX.12.040501}.
Spin control can rely on different approaches, such as the application of an external magnetic field~\cite{PhysRevApplied.14.014004}, the injection of electrical current~\cite{RevModPhys.91.035004, bodnar2018,Quessab2018}, the application of mechanical strain~\cite{bhattacharya2019review}, and the application of laser pulses~\cite{PhysRevB.73.014421, RevModPhys.82.2731, PhysRevLett.99.047601, kimel2005ultrafast, grigorev2022optically}.
Regardless of the means of controlling the magnetic state, one of the key aspects in building any technology based on magnetic materials is the ability to tailor the strength of the magnetic anisotropy.
Different approaches have already been established to control magnetic anisotropy, particularly in ferromagnetic films with perpendicular magnetic anisotropy (PMA).
They range from selecting material composition to controlling layer architecture  ~\cite{Aman2026} and sample geometry~\cite{huang1999copt, zeng2002copt} to introducing order or disorder through ion bombardment~\cite{ravelosona1999ion, fassbender2004magnetic, MAZALSKI2019317}.

Laser light has also been efficiently used to control magnetic anisotropy~\cite{afanasiev2021control, riddiford2025laser}.
For instance, early studies on Pt/Co/Pt systems demonstrated that laser irradiation can locally modify magnetic anisotropy and trigger domain structure transformations~\cite{Kisielewski2008,Kisielewski2014}.
For example, Talapatra et al have shown that laser pulses applied to a Co/Pt multilayered system  can induce localized magnetic softening, characterized by a selective $50\%$ reduction in the anisotropy energy, thereby rendering the irradiated zones magnetically softer and causing the equilibrium domains to shrink~\cite{TalapatraCoPt}.
Beyond uniform tuning, recent advances highlight the immense potential of engineering local variations in material properties at the micrometer scale.
Specifically, it has been demonstrated that creating complex two-dimensional (2D) patterns with continuous gradients in magnetic anisotropy, interlayer exchange coupling, and ferrimagnetic compensation can unlock ground-breaking functionalities, such as spin-wave bandpass-filters and architectures for passive domain wall resetting~\cite{riddiford2025laser}. 
Previous studies have established helicity-dependent domain-wall motion and laser-driven transformations of magnetic domain patterns in Co/Pt multilayers~\cite{Novakovic2020}.
However, the spatially resolved irreversible enhancement of effective anisotropy and coercivity following localized high-fluence laser annealing, together with the correlation with equilibrium domain size, remain  unexplored.

In this work, we investigate the impact of laser irradiation on perpendicularly magnetized Co/Pt multilayers across two distinct fluence regimes by employing a complementary combination of vibrating sample magnetometry (VSM), wide-field Kerr microscopy and magnetic force microscopy (MFM).
This multi-scale approach reveals a transition from non-destructive spin manipulation to permanent magnetic landscape engineering.
At low fluences ($1.2\,\text{mJ/cm}^2$), wide-field Kerr microscopy demonstrates a non-invasive, helicity-dependent accumulation of magnetic phases (i.e., preferential growth of domains with matching magnetization orientation) driven by laser-induced spin torques assisted by film heating, thereby preserving the original parameters of the film.
Conversely, at higher fluences ($5\,\text{mJ/cm}^2$), the laser exposure acts as localized thermal annealing, permanently hardening the magnetic system. 
Macroscopic VSM measurements alongside nanoscale MFM and magneto-optical imaging reveal a substantial enhancement of the intrinsic perpendicular magnetic anisotropy (PMA) and a twofold increase in the local coercive field ($H_\mathrm{c}: 18\,\text{mT} \rightarrow 36\,\text{mT}$), while maintaining saturation magnetization and excellent loop squareness ($M_\mathrm{r}/M_\mathrm{s} \approx 1$).

This dual-regime behavior and the subsequent magnetic hardening fundamentally distinguish our study from the laser-induced softening reported in Ref.~\cite{TalapatraCoPt}.
In contrast to anisotropy degradation and domain shrinking, our work demonstrates a localized laser-induced enhancement of the intrinsic perpendicular anisotropy and drives a sizable expansion of the equilibrium domain width from $0.9\,\mu\text{m}$ to $2.2\,\mu\text{m}$.
By utilizing spatially resolved coercivity mapping to reveal a localized high-fluence enhancement of intrinsic PMA, alongside a reversible low-fluence regime, this work moves beyond conventional laser-induced demagnetization. It provides a versatile pathway for precisely tuning coercive fields and engineering continuous 2D magnetic landscapes.

\section{Experimental}

Experiments on optically-induced magnetization and anisotropy modification were conducted on a Co/Pt multilayer with large PMA~\cite{Kisielewski2014}.
The PMA thin film was prepared by DC magnetron sputtering at room temperature with the following layer architecture: 
Si\slash SiO2(300)\slash Ta(5)\slash Pt(2)\slash [Co(0.4)/Pt(0.9)]$_4$\slash Co(0.4)\slash Pt(2).
The thickness of each layer (numbers in brackets) are given in nanometers.

Magneto-optical Kerr effect (MOKE) microscopy was performed using a custom-built wide-field Kerr microscope setup.
Out-of-plane magnetic fields up to $\mu_0 H = 100$\, mT were applied along the optical axis (perpendicular to the sample surface) and real-time image processing was used to maximize the magneto-optical contrast.
Depending on the required field of view, either a $20\times$ (numerical aperture $\text{NA} = 0.4$) or a $50\times$ ($\text{NA} = 0.8$) objective lens was employed.

Spatially resolved magnetization curves were reconstructed by dividing the imaged area into a matrix of square elements and plotting the intensities of these elements ~\cite{SoldatovHcLoops2026}.
For this purpose, neighboring pixels were binned post-acquisition into $2\times2$ blocks acting as virtual pixels and are designated as regions of interest (ROIs) in the text.
The magnetization reversal curves were obtained using a quasistatic domain imaging approach, in which the magnetic field was varied in 1000 discrete steps with a measurement time of approximately 1 second per point and averaging of 16 frames~\cite{SoldatovFaraday2017,SoldatovKerrSepar2017}.
Vibrating sample magnetometry (VSM, Quantum Design Versalab) was used to perform the global magnetization measurements shown in Fig.~\ref{LoopsGLobal}.
During hysteresis measurements, the magnetic field was swept at the rate of 2.5 mT/sec, and data were sampled every 1 sec.
Magnetic force microscopy (MFM) measurements were conducted with a Bruker Icon system in standard lift mode.
To avoid manipulation (movement) of domain walls by the scanning tip, custom-made MFM probes with low coating thickness (< 12 nm) of a Co-based magnetic layer were applied.
They allowed stable imaging of the labyrinth domain patterns in the laser exposed regions at a lift height of 80 nm.
In the pristine, ac-demagnetized sample regions a lift height of 80 nm leads to visible modifications of the domain wall positions, hence a larger lift height of 100 nm has been applied.
This observation suggests a slight, albeit macroscopically unresolvable, increase in local coercivity induced by laser exposure.

As the source of laser radiation, we used a titanium-sapphire regenerative amplifier system (Astrella, Coherent Inc.), which provided fundamental laser light at 800 nm wavelength.
To generate the second harmonic, the fundamental beam was focused into a beta-barium borate ($\beta$-BBO) nonlinear crystal. After frequency doubling, the resulting laser beam had a central wavelength of 400 nm, operated at a 1 kHz repetition rate, and had an approximate pulse duration of 100 fs.

\section{Helicity-dependednt domain manipulation}
\label{OpticalSwitching}

\begin{figure}
\center
\includegraphics[width=1\linewidth,clip=true]{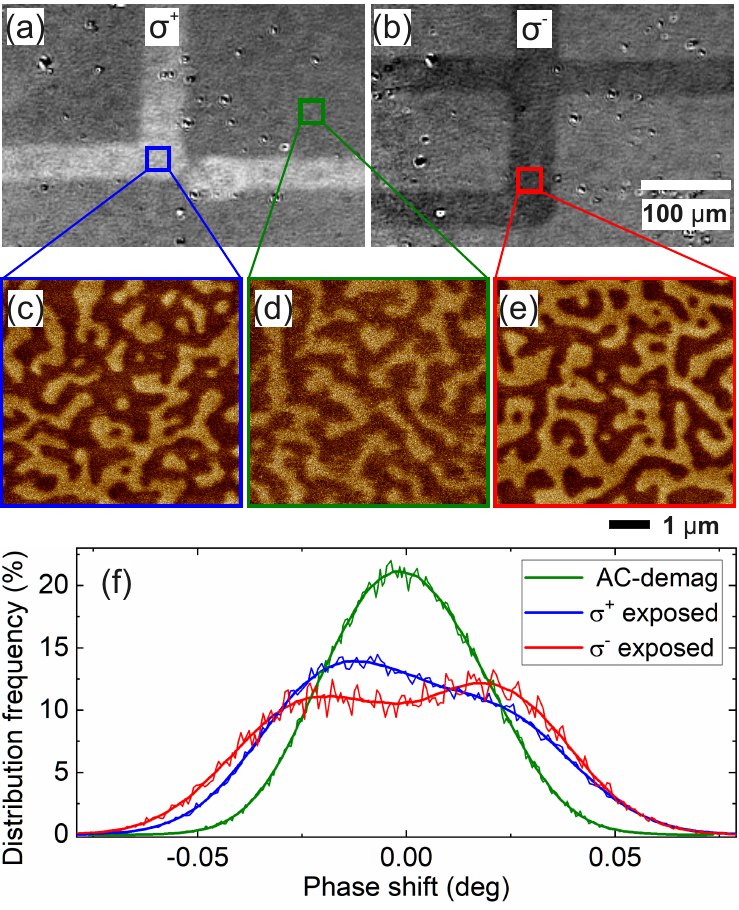}
\caption{Helicity-dependent laser modification of the magnetic domains in Co/Pt thin film by laser with fluence of $1.2\text{ mJ/cm}^2$.
Wide-field MOKE images revealing a \textbf{(a)} bright track induced by left-handed ($\sigma^+$) and a \textbf{(b)} dark track induced by right-handed ($\sigma^-$) circularly polarized laser light.
MFM images of the domain morphology in the \textbf{(c)} unexposed AC-demagnetized, \textbf{(d)} $\sigma^+$-modified, and \textbf{(e)} $\sigma^-$-modified states. \textbf{(f)} Corresponding MFM phase-shift histograms, demonstrating the transition from a symmetric domain area distribution to a helicity-dependent asymmetry after the laser treatment.
}\label{CoPt_Fig1}
\end{figure}

To investigate the impact of laser irradiation on the film magnetization, the sample was initially demagnetized using an AC magnetic field, yielding a characteristic labyrinth domain pattern (Fig.~\ref{CoPt_Fig1}c).
MFM data analysis confirms an equal distribution of up- and down-magnetized domains in which the phase shift distribution of both domain polarities merges indistinguishably into one symmetric phase shift histogram (green curve in Fig.~\ref{CoPt_Fig1}f).
Subsequently, the pulsed laser beam with photon energy of 3.09 eV was focused onto the sample surface into a $120\,\mu\text{m}$ spot with a peak fluence of $1.2\text{ mJ/cm}^2$ and scanned across the sample to form continuous lines.
As seen in Fig.~\ref{CoPt_Fig1}a and b, the laser exposure induces either bright or dark tracks against the unexposed gray matrix depending on the beam helicity, corresponding to areas scanned by left-handed ($\sigma^+$) or right-handed ($\sigma^-$) circularly polarized light, respectively.

A closer inspection of the magnetic domain structure using MFM (Fig.~\ref{CoPt_Fig1}c-e) reveals that although the overall domain morphology remains unchanged compared to the unexposed regions, the area ratio of the up- and down-magnetized states deviates from equilibrium.
For positive helicity $\sigma^+$ (blue curve), the right MFM histogram peak (negative phase shift) dominates over the left peak (positive phase shift), whereas for negative helicity $\sigma^-$ (red curve), this asymmetry is inverted.
Such helicity-dependent domain redistribution is consistent with helicity-dependent domain-wall motion driven by the combined effects of laser heating, temperature gradients, domain-wall pinning, and helicity-dependent optical excitation\cite{PhysRevLett.99.047601}.
This subtle deviation from the strict 50/50 domain area ratio is sufficient to break the almost-perfect local cancellation of stray fields inherent to the symmetric labyrinth.
Crucially, local magneto-optical hysteresis loops measured within the laser tracks revealed no measurable changes in the coercive field or loop shape.
These magnetic measurements indicate no structural modifications or interdiffusional degradation in the Co/Pt film within our detection limits at such low fluence, consistent with a reversible laser-induced magnetic reorientation without detectable permanent modification of the quasistatic magnetic properties.
Applying a saturating out-of-plane magnetic field uniformly erases the optically induced domains simultaneously with the surrounding matrix, confirming that this low-fluence regime acts purely as a reversible magnetic write-process without permanently engineering the local anisotropy.

\section{Modification of the local anisotropy}
\label{StructuralModif}

\begin{figure}
\center
\includegraphics[width=1\linewidth,clip=true]{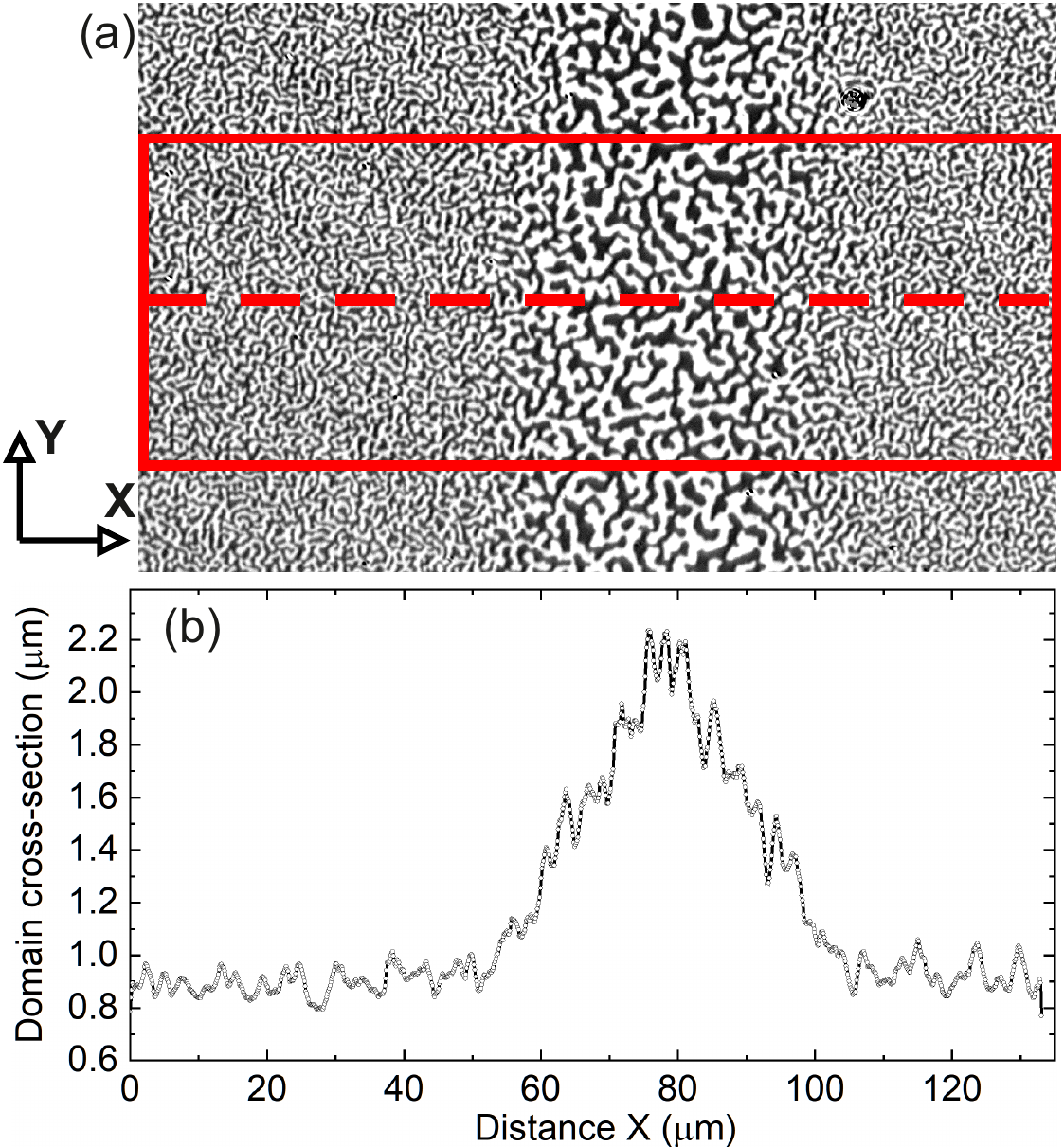}
\caption{Modification of the magnetic domain structure in the Co/Pt film under high-fluence laser irradiation ($5\text{ mJ/cm}^2$). \textbf{(a)} MOKE microscopy image of the equilibrium domain pattern in the AC-demagnetized state, showing a pronounced increase in the average domain size within the laser track compared to the pristine area. \textbf{(b)} Average domain cross-section along the Y-direction plotted as a function of the lateral coordinate X across the laser line.
}\label{CoPt_DomainBurn}
\end{figure}

While low laser fluences (such as $1.2\text{ mJ/cm}^2$) leave the intrinsic and extrinsic magnetic properties unaltered — inducing a purely local reorientation of magnetization — higher fluences trigger permanent modifications due to structural changes.
To demonstrate this, the sample surface was scanned at an increased pulse fluence of $5\text{ mJ/cm}^2$ with the same photon energy of 3.09 eV.
The resulting modifications were evaluated by analyzing the equilibrium multi-domain patterns obtained after AC-demagnetization in a perpendicular magnetic field.
In ultrathin films, the equilibrium domain width is governed by a delicate balance between the magnetostatic energy ($\propto M_s^2$) and the domain wall energy ($\propto \sqrt{A K_u}$), where A is the exchange stiffness  constant and $K_u$ is the uniaxial magnetic anisotropy constant~\cite{Hubert1998}.  
As shown in Fig.~\ref{CoPt_DomainBurn}a, the domain size within the laser-exposed track is significantly larger than in the surrounding pristine matrix.

Quantitatively, the average domain cross-section along the Y-direction increases more than twofold, expanding from $0.9\,\mu\text{m}$ in the unexposed region to a peak value of $2.2\,\mu\text{m}$ at the center of the laser track (Fig.~\ref{CoPt_DomainBurn}b).
The approximately Gaussian spatial dependence of the domain size suggests a close relation between the local irradiation dose and the resulting magnetic modification.

\begin{figure}
\center
\includegraphics[width=1\linewidth,clip=true]{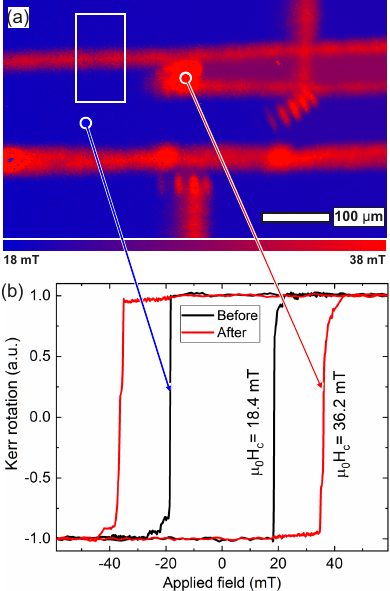}
\caption{Spatial mapping of the magnetic properties after high-fluence laser modification. \textbf{(a)} Spatial distribution of the local coercive field $H_\mathrm{c}$ extracted from pixel-by-pixel MOKE loops. 
The white rectangle indicates the region where the equilibrium magnetic domains were extracted for Fig.~\ref{CoPt_DomainBurn}.
\textbf{(b)} Representative local MOKE hysteresis loops measured in the pristine matrix (blue curve, $H_\mathrm{c} = 18.4$\,mT) and at the center of the laser track (red curve, $H_\mathrm{c} = 36.2$\,mT).
}\label{CoPt_HcDistr}
\end{figure}

To quantify changes in the magnetic parameters, the out-of-plane magnetic field $\mu_0 H$ was cycled between $-100$\,mT and $+100$\,mT to obtain position-dependent magnetization reversal curves.
These local hysteresis loops were obtained by tracking the image grey level averaged over $2\times2$ pixel areas, from which the local coercive field $H_\mathrm{c}$ was subsequently extracted.
Figure~\ref{CoPt_HcDistr}a maps the resulting spatial distribution of $H_\mathrm{c}$ across the sample surface.
In the pristine, unexposed matrix, the film exhibits a coercive field of approximately $18$\,mT (blue loop in Fig.~\ref{CoPt_HcDistr}b).
In contrast, the laser-irradiated tracks undergo significant magnetic hardening, with $H_\mathrm{c}$ practically doubling to a peak value of $\approx36$\,mT at the track center (red loop in Fig.~\ref{CoPt_HcDistr}b).
The observed increase in coercivity, together with the persistent modification of the domain structure and magnetic anisotropy, demonstrates that the laser exposure induces an irreversible modification of the magnetic properties. The microscopic origin of this modification may involve changes in interface structure, strain relaxation, defect redistribution, and/or local interdiffusion.
In addition, localized peaks in the $H_\mathrm{c}$ distribution (Fig.~\ref{CoPt_HcDistr}a) are consistent with regions that received a higher cumulative irradiation dose during scanning.
This manifestation of a cumulative thermal effect underscores that the tuning method is highly sensitive to the local irradiation dose, enabling precise grayscale control over the magnetic landscape.

\begin{figure}
\center
\includegraphics[width=1\linewidth,clip=true]{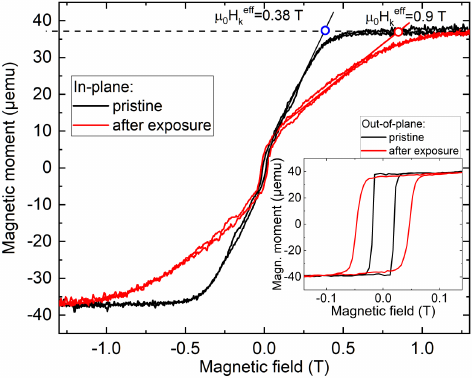}
\caption{Integral magnetization reversal curves measured for a $3\times3.5\,\text{mm}^2$ sample before (black) and after (red) laser exposure with the magnetic field applied within the sample plane. The inset displays the corresponding macroscopic hysteresis loops obtained in an out-of-plane magnetic field.
}\label{LoopsGLobal}
\end{figure}

To understand the driving forces behind the observed domain expansion, several physical mechanisms must be considered: 
(i)~a reduction in domain wall pinning sites density~\cite{bertotti1998hysteresis}, 
(ii)~a decrease in $M_\mathrm{s}$, which lowers the magnetostatic driving force for domain splitting~\cite{Hubert1998}, or 
(iii)~an increase in $K_\mathrm{u}$, which raises the energy cost of domain wall formation ($\sigma_\mathrm{w} = 4\sqrt{A K_\mathrm{u}}$)~\cite{skomski2008simple}.

The first scenario involving decreased pinning can be confidently ruled out by our localized MOKE measurements (Fig.~\ref{CoPt_HcDistr}); a reduction in the density of pinning sites would inevitably lower the coercive field, whereas we observe a prominent twofold increase in $H_\mathrm{c}$ within the areas modifed by laser pulses.

To differentiate between the remaining options and evaluate changes in the fundamental intrinsic parameters ($M_\mathrm{s}$ and $K_\mathrm{u}$), global magnetization measurements were performed.
A smaller section of the sample ($3\times3.5\,\text{mm}^2$) was prepared to measure the absolute magnetic moment before and after the entire area was uniformly raster-scanned with high-fluence laser exposure ($5\text{ mJ/cm}^2$).
As shown in Fig.~\ref{LoopsGLobal}, the saturation magnetic moment in both the in-plane (main panel) and out-of-plane (inset) configurations remains perfectly identical before and after laser irradiation.
The unchanged room-temperature saturation magnetization rules out a significant change in $M_\mathrm{s}$ as the primary origin of the observed domain expansion, implying that both the Curie temperature and the core electronic structure should remain unaltered — the exchange stiffness constant $A$ can be safely assumed to be constant as well.

Crucially, the in-plane (hard-axis) magnetization curve becomes significantly more slanted after exposure, requiring a much higher external field to achieve saturation.
This macroscopic confirmation of a substantial increase in the effective anisotropy field ($B_\mathrm{k}^\mathrm{eff}$), provides direct experimental evidence that the laser treatment enhances the intrinsic PMA.
Quantitatively, evaluating the effective magnetic anisotropy $K_{\mathrm{eff}} = \frac{1}{2} M_{\mathrm{s}} \mu_0 H_{\mathrm{k}}^{\mathrm{eff}}$ using the effective anisotropy field $\mu_0 H_{\mathrm{k}}^{\mathrm{eff}}$ determined from the slope of the hard-axis VSM magnetization curves (assuming a stack-averaged $J_{\mathrm{s}} = 0.68~\mathrm{T}$) yields a more than twofold increase from $1.03 \times 10^5~\mathrm{J/m}^3$ ($103~\mathrm{kJ/m}^3$) in the pristine film to $2.44 \times 10^5~\mathrm{J/m}^3$ ($244~\mathrm{kJ/m}^3$) after the $5\text{ mJ/cm}^2$ laser treatment.

Consistent with this enhanced anisotropy, we observe an increase in the coercive field in the out-of-plane configuration from 18~mT up to 36~mT, similar to the local observations in the previous section.
Notably, despite this significant magnetic hardening, the out-of-plane loops (Fig.~\ref{LoopsGLobal}, inset and Fig.~\ref{CoPt_HcDistr}b) retain their high squareness ($M_\mathrm{r}/M_\mathrm{s} \approx 1$) and sharp switching behavior, indicating that the collective character of the reversal is largely preserved\cite{Quessab2018}.
To evaluate whether this modification introduces extrinsic defect-induced pinning, we calculated the effective Kronmüller coefficient, defined via $K_{\mathrm{eff}}$ as $\alpha_{\mathrm{eff}} = \mu_0 H_{\mathrm{c}} / \mu_0 H_{\mathrm{k}}^{\mathrm{eff}}$ \cite{skomski2008simple}.
The calculated parameter remains virtually constant, shifting only slightly from $\alpha_{\mathrm{eff}} \approx 0.048$ (pristine) to $0.040$ (high fluence).
The approximately unchanged ratio \(H_\mathrm{c}/\mu_0 H_\mathrm{k}^\mathrm{eff}\) (stability of $\alpha_{\mathrm{eff}}$) suggests that the relative relationship between coercivity and the effective anisotropy field is preserved after irradiation.
This observation is consistent with the increase in coercivity primarily being attributed to enhanced anisotropy, although changes in the detailed pinning landscape cannot be entirely excluded.

Integrating these multi-scale observations allows us to construct a consistent physical picture of the laser-induced modification in our Co/Pt films.
While conventional thermal annealing of Co/Pt multilayers often induces interdiffusion and degrades perpendicular anisotropy, short laser-induced thermal pulses operate in a localized relaxation regime that preserves interfacial integrity \cite{Medapalli2017}.
On the intrinsic level, this thermal energy may facilitate the relaxation of growth-induced structural stresses and sharpens the interfacial texturing of the Co/Pt multilayers without triggering destructive chemical mixing.
This manifests macroscopically as an enhanced PMA and coercivity in global hysteresis measurements (Fig.~\ref{LoopsGLobal}) and microscopically as a doubling of the equilibrium domain width (Fig.~\ref{CoPt_DomainBurn}) due to the increased cost of domain walls, combined with a doubling of the local coercivity.
Consequently, this high-fluence regime provides an efficient route for ``magnetic landscape engineering,'' where the local magnetic profile is permanently tailored primarily via intrinsic anisotropy tuning \cite{TalapatraCoPt}.

\section*{Conclusions}
\label{ConclusionChapter}

We have demonstrated dual-regime laser control over both the transient magnetic states and permanent intrinsic properties of perpendicularly magnetized Co/Pt multilayers using a complementary combination of wide-field Kerr microscopy, MFM, and integral magnetometry. 

In the low-fluence regime ($1.2\,\text{mJ/cm}^2$), a non-invasive, helicity-dependent accumulation of magnetic phases is achieved.
A modest external magnetic field can fully erase this optically induced magnetization state, indicating that the magnetic properties of the sample, such as local magnetic anisotropy and coercivity, remain unaffected.

Conversely, the high-fluence regime ($5\,\text{mJ/cm}^2$) induces permanent magnetic hardening via a localized laser-annealing process.
Multi-scale analysis reveals a substantial increase in the effective perpendicular anisotropy, manifested macroscopically as a doubling of the in-plane saturation field.
Spatially resolved MOKE loop mapping across the irradiated tracks enabled us to reconstruct the local coercivity distribution, revealing a sharp, prominent increase in the out-of-plane coercive field while preserving excellent loop squareness and collective reversal dynamics.
This enhancement of the perpendicular magnetic anisotropy directly governs the magnetic texture, driving more than twofold expansion of the equilibrium stripe domain width from $0.9\,\mu\text{m}$ to $2.2\,\mu\text{m}$ as confirmed by high-resolution imaging.

Our findings bridge the gap between optical spin manipulation and deterministic, local anisotropy tuning. The ability to map and control both the local coercivity distribution and the equilibrium domain size opens an efficient pathway toward the realization of lithography-free spatial engineering of magnetic landscapes for advanced topological and texture-based spintronic logical elements.

\section*{Acknowledgments}
\label{AcknowledgmentAll}
This work was supported by the Bundesministerium für Forschung, Technologie und Raumfahrt (BMFTR) under Grant No. 05K25BD1 (I.S.), and by the Leibniz Association within the framework of the nanoALT project, funded through the Leibniz Competition (S.B.) and by the German Research Foundation (DFG) via project 43356004 (V.N.).

\section*{Data availability}
\label{DataExist}
All data generated or analysed during this study are included in this manuscript.

\section*{Authors contributions}
\label{Contributions}
Ivan Soldatov primarily wrote the manuscript. All authors contributed to manuscript editing.
Stanislav Bodnar and Ivan Soldatov conducted magneto-optical Kerr effect (MOKE) measurements.
Volker Neu performed magnetic force microscopy (MFM) measurements, and Aman Singh conducted vibrating sample magnetometry (VSM) measurements.
Yana Vaynzof and Andrei Scherstobitov discussed the scientific results.

\bibliographystyle{naturemag}

\bibliography{References}

\end{document}